# Human Capital, Not Model Benchmarks, Predicts Hybrid Intelligence in Forecasting

Vivienne Ming
The Human Trust; Possibility Science; UCL Global Business School for Health

*Preprint — pilot study. Findings are preliminary and intended to motivate a pre-registered replication.*

**Significance.** Whether pairing people with AI helps or hurts is usually reported as a single average effect. Using a real-money prediction market (Polymarket) as an objective, externally resolved benchmark, this pilot shows that the value of human–AI collaboration depends on a specific, measurable form of human capital. Analyzed at the level of the individual forecaster, hybrid performance is trimodal: most people either deferred to the model (matching it) or used it to rubber-stamp a prior guess (performing worse than the model alone), while a minority engaged in genuine complementary reasoning and reached accuracy matching or even exceeding (i.e., lower error than) the market itself. Collaborative traits—perspective-taking, intellectual humility, and curiosity—rather than raw cognitive ability or model benchmarks, distinguished who reached that mode. The results are preliminary but statistically robust, and motivate a pre-registered replication now in preparation.

## Introduction

Studies of human–AI collaboration report conflicting effects: some find augmentation of lower-skilled workers (1, 2), others find the benefits accrue mainly to experienced users (3), still others find that access to capable models degrades human performance relative to the model working alone (4, 5), and others show substantial returns on complex, elite tasks (6). A recurring confound is evaluation. In some cases AI systems may simply have encoded the problem and its solution, so that measured "ability" reflects contamination rather than reasoning (7, 8). And even on less constrained tasks, "quality" is often scored by expert raters who are themselves swayed by the fluency of AI-assisted prose rather than the substance of the work (9, 10). Forecasting against a real-money prediction market sidesteps both problems: every prediction is scored against an externally resolved ground truth on a common, style-free metric.

I report a pilot designed to estimate (i) whether human capital moderates the value of AI assistance in forecasting, and (ii) whether distinct interaction styles mediate that moderation. Because each participant entered their own probabilistic forecast, all analyses are conducted at the level of the individual forecaster. I frame the work as hypothesis-generating: several cells are small, and the aim is to obtain effect-size estimates—now statistically supported—to power a confirmatory trial in preparation.

## Methods

**Participants and design.** One hundred eight adults were recruited by flyer in Berkeley, CA and compensated $20 per session for three sessions. The main study comprised 78 participants (42 UC Berkeley students; 36 community adults; mean age 28.7, range 18–60) organized into 26 three-person teams and assigned either to a Human-only condition (12 teams) or a Hybrid condition (14 teams) with access to one of four large language models. A separate pool of 30 participants (10 teams) worked with a "Socratic" model that withholds answers, reported separately (Supplementary S1). Participants collaborated in teams but each entered an individual forecast; analyses are at the

participant level. One participant was excluded for an out-of-range Brier score (−0.005), leaving 77 analyzed in the main pool.

**Forecasting task and ground truth.** The 30 questions were live Polymarket contracts resolving between November 2025 and January 2026, spanning economics, international relations, and business. Each question's externally certified resolution ($o \in \{0,1\}$) provided ground truth, and each team forecast 10 questions drawn at random from the pool. The four models also forecast all 30 questions independently (AI-only baseline): Llama 3.1 (8B), Qwen3 (8B), GPT-4o, and Gemini 3 Pro. As an external reference I recorded the market's implied probability (scaled Brier 3.5).

**Outcome metric.** Forecast accuracy is the Brier score, the mean squared error of the probabilistic forecast, $B = (1/N)\ \Sigma\ (p_i - o_i)^2$, reported scaled ×100 (range 0–100; lower is better; an uninformative 0.5 forecast scores 25).

**Human-capital measures.** Session 1 administered Raven's Advanced Progressive Matrices Set II (fluid reasoning), the ICAR public-domain g proxy, the IRI Perspective-Taking subscale, the Comprehensive Intellectual Humility Scale, an epistemic-curiosity scale, the Brief Resilience Scale, and the TIPI Big-Five inventory.

**Interaction styles.** Hybrid participants were classified post hoc from their process into three emergent styles: Automators (adopt the model's answer with little added reasoning), Validators (use the model to check a prior guess), and Cyborgs (iterative complementary reasoning). Because style is emergent rather than assigned, it is collinear with human capital; comparisons across styles are descriptive.

**Analysis.** I report participant-level condition means with 95% confidence intervals and individual points (Fig. 1). Given heterogeneous within-condition variances I use a non-parametric omnibus test (Kruskal–Wallis) with Welch's t-tests for post-hoc pairwise comparisons, and Pearson correlations of human capital with accuracy. Effect sizes are Cohen's d.

## Results

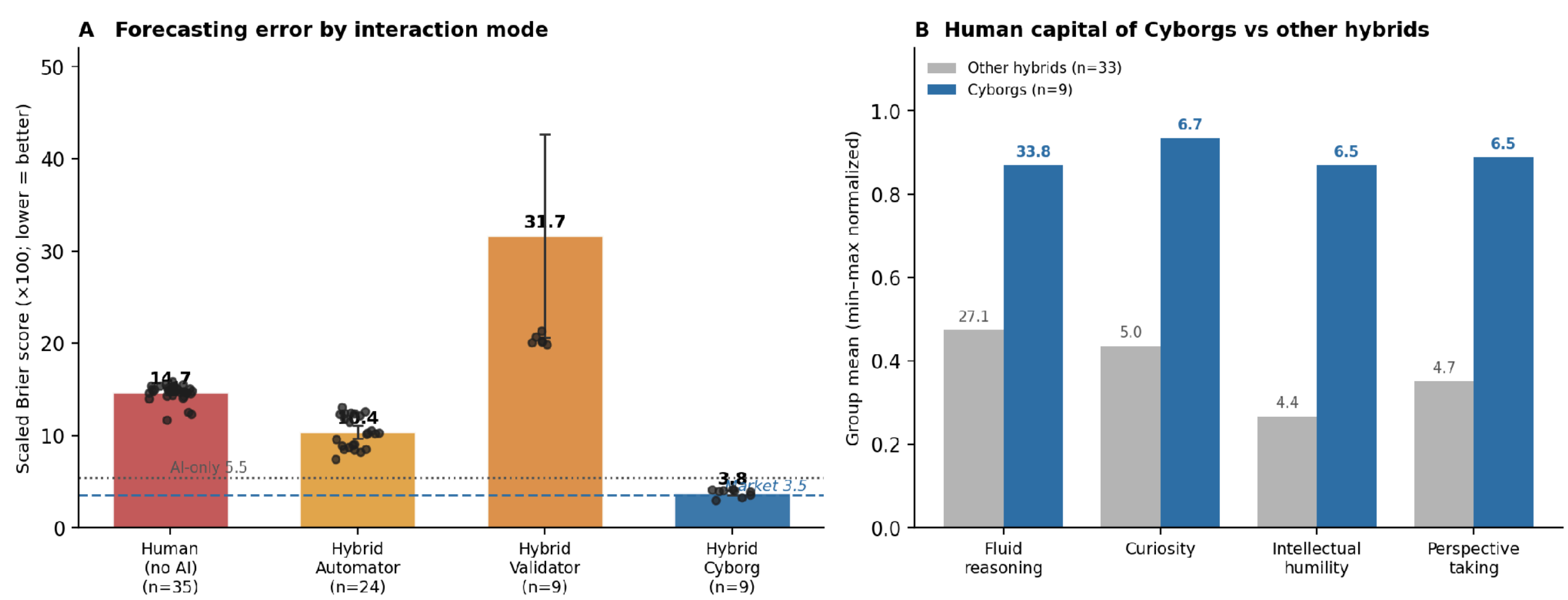

**Figure 1. Participant-level performance and the human capital of Cyborgs.** (A) Scaled Brier score (×100; lower is better) by condition; bars are participant means, dots individual forecasters, whiskers 95% CIs. Dashed line, Polymarket benchmark (3.5); dotted line, mean AI-only baseline (5.5). (B) Cyborgs versus other hybrid forecasters on four human-capital traits (group means, min–max normalized; raw means printed). The Socratic pool is reported separately (Supplementary S1); one participant was excluded for an out-of-range score (see Limitations).

**A trimodal "K-shape."** Forecasting error differed sharply across conditions (Kruskal–Wallis $\chi^2$ across the four groups, $p < 10^{-13}$; Fig. 1A). Human-only forecasters scored 14.7 (95% CI 14.4–15.0, $n = 35$), well above chance but far behind the models. Crucially, hybridization did not uniformly help. Automators reproduced roughly their tool's behavior but not its accuracy (10.4, CI 9.7–11.0, $n = 24$)—better than humans alone ($p < .001$) yet worse than the AI-only baseline (5.5). Validators—the classic "human-in-the-loop" who uses AI to check a prior guess—were significantly worse than humans working alone (31.7, CI 20.6–42.7, $n = 9$; vs. human-only $p = .017$). Only Cyborgs improved on the machine: 3.8 (CI 3.5–4.1, $n = 9$), lower error than every individual AI model and statistically indistinguishable from—indeed numerically below the mean of—the strongest models, significantly better than Automators ($p < .001$).

**Relation to the market.** Cyborg forecasters matched or exceeded (i.e., achieved lower error than) the market benchmark of 3.5, whereas Automators and Validators did not approach it. The comparison of Cyborgs to the pooled AI-only baseline was in the same direction but only marginal ($p = .06$), reflecting the four-model baseline rather than a weak effect; Cyborg error was below each of the four models individually.

**Human capital predicts hybrid accuracy—benchmarks do not.** Among the four models, AI-only accuracy tracked scale and benchmark standing (GPT-4o and Gemini 3 Pro, 4.6; Qwen3, 5.5; Llama 3.1, 7.1). Yet among hybrid forecasters, cognitive ability—the human analog of a benchmark score—did not predict accuracy (g, $r = -0.10$; fluid reasoning, $r = -0.08$; both ns). Collaborative human capital did: perspective-taking predicted lower error ($r = -0.32$, $p = .04$), with curiosity ($r = -0.26$) and intellectual humility ($r = -0.25$) trending. The contrast with solo human forecasting is striking: among Human-only participants, every trait strongly predicted accuracy—intellectual humility ($r = -0.66$, $p < .001$), and g, fluid reasoning, curiosity, and perspective-taking all near $r = -0.45$ ($p < .01$). Access to AI thus decouples raw ability from outcome—the model supplies a floor—while collaborative capacity instead governs who transcends that floor.

**Who becomes a Cyborg.** Human capital sharply gated emergence of the Cyborg mode. Cyborgs exceeded other hybrid forecasters on intellectual humility (6.5 vs. 4.4, $d = 4.1$), curiosity (6.7 vs. 5.1, $d = 4.0$), and perspective-taking (6.5 vs. 4.6, $d = 3.2$), with smaller but reliable differences in fluid reasoning ($d = 1.5$) and g ($d = 1.7$) (all $p < .01$; Fig. 1B). The separation was strong enough that a logistic model of Cyborg emergence on these traits was not identifiable at this sample size—evidence of a very strong association rather than a null.

**Qualitative complementarity.** Process notes suggest a division of labour in Cyborg pairs: humans drove exploration of the ill-posed, contingent parts of a question (what could intervene, what the market might be missing) while the model anchored the well-posed, data-bound parts. Automators outsourced both; Validators used the model to crystallize, rather than challenge, their own reasoning.

## Discussion

These results reframe the question of AI and productivity from whether AI helps to when and how it helps. Pooling all hybrid interaction styles (Automators, Validators, and Cyborgs) into a single average "hybrid" cell hides a strongly trimodal outcome—including a condition (Validator) that underperforms the AI alone and even the unaided human. The moderator that organizes the pattern is collaborative human capital, especially perspective-taking, curiosity, and intellectual humility, consistent with broader findings that heterogeneity in collaboration ability drives the returns to hybrid intelligence (11–14). Notably, the human capital that matters is not the human analog of a model benchmark: raw cognitive ability predicted solo human accuracy and tracked the models' own benchmark ordering, but did not predict who succeeded with AI. What makes a model score well, or a person reason well alone, is not what makes a human–AI team effective.

**Limitations.** This is a pilot and should be read as such. The Cyborg and Validator cells are small (n = 9 each); interaction style is emergent and collinear with human capital, so style comparisons are descriptive, not causal; and within-condition outcome variance is low, which inflates some standardized effect sizes—interpretation here rests on the rank-based omnibus test and confidence intervals rather than on d. The market comparison is limited by a four-model baseline and per-question calibration was not independently re-audited. A pre-registered, adequately powered replication (in preparation) will (i) pre-register the style taxonomy and a human-capital composite, (ii) manipulate rather than merely measure interaction style where feasible, and (iii) fix the forecast-elicitation timestamp relative to market state.

## Supplementary notes (incidental, not part of the pre-specified design)

**S1. A "Socratic" model that never answers.** A separate pool of 30 participants (10 teams) worked with the lowest-tier open model re-tuned to withhold direct answers and instead prompt reasoning. It failed every accuracy benchmark, yet shifted behavior in the predicted direction: the Cyborg rate rose to 30% (vs. 21% in the main hybrid pool), and would-be Automators, unable to copy an answer, fell to human-level accuracy (14.9 vs. 10.4 in the main pool). Cyborgs remained accurate (4.2). User satisfaction, however, was uniformly poor (post-task CSAT ≈ 1–2 on a 10-point scale).

**S2. EEG cognitive engagement.** Midway through data collection, a mobile EEG (UC Berkeley Neurotech Collider Lab) recorded relative gamma power as a proxy for cognitive effort in volunteers. Automators showed ≈43% lower relative gamma than human-only participants (0.57 vs. 1.0). This is an opportunistic, unblinded observation in 7 participants, with gamma-as-engagement a contested operationalization; it is reported only to motivate instrumented follow-up.